\documentclass{desyproc}

\newcommand{\aap}{Astron. Astrophys.}

\newcommand{\araa}{Ann. Rev. Astron. Astrophys.}

\newcommand{\jcap}{JCAP}

\newcommand{\prd}{Phys. Rev. D}
\newcommand{\apj}{Astrophys. J.}
\begin{document}
%------------------------------------
\title{The effect of photon-axion-like particle conversions in galaxy clusters on very high energy $\gamma$-ray spectra}

%for single authors the superscripts are optional
\author{{\slshape Manuel Meyer$^1$, Dieter Horns$^{1}$, Luca Maccione$^2$, Alessandro Mirizzi$^3$, Daniele Montanino$^4$ and Marco Roncadelli$^5$}\\[1ex]
$^1$ Institut f\"ur Experimentalphysik, Universit\"at Hamburg, Luruper Chaussee 149, 22761 Hamburg, Germany\\
$^2$ Arnold Sommerfeld Center, Ludwig-Maximilians-Universit\"at, Theresienstraße 37, 80333 Munich, Germany; Max-Planck-Institut f\"ur Physik (Werner-Heisenberg Institut), F\"ohringer Ring 6, 80805 Munich, Germany\\
$^3$ II Institut f\"ur Theoretische Physik, Universit\"at Hamburg, Luruper Chaussee 149, 22761 Hamburg, Germany\\
$^4$ Dipartimento di Matematica e Fisica ``Ennio de Giorgi'', Universit\`a del Salento and Sezione INFN di Lecce, Via Arnesano, I-73100 Lecce, Italy\\
$^5$ INFN, Sezione di Pavia, Via A. Bassi 6, 27100 Pavia, Italy
}

% if the proceedings are available online (e.g. at Indico)
% please enter the contribution ID or file_name below for the DOI
%\contribID{32}
%\contribID{familyname\_firstname}

% TO THE CONFERENCE EDITORS: 
% please update the following information      
% before sending the template to the authors
% \confID{800}  % if the conference is on Indico uncomment this line
\desyproc{DESY-PROC-2012-04}
\acronym{Patras 2012} % if you want the Acronym in the page footer uncomment this line
\doi  % if there is an online version we will register DOIs

\maketitle

\begin{abstract}
Very high energy (VHE, energy $\gtrsim 100\,$GeV) $\gamma$-rays originating from extragalactic sources interact with low energy photons of background radiation fields and 
produce electron-positron pairs. Alternatively, in the presence of ambient magnetic fields, they can convert into hypothetical axion-like particles (ALPs), 
pseudo-scalar spin-0 bosons, predicted by extensions of the standard model.
These particles propagate unimpeded over cosmological distances. 
Here, the effect of photon-ALP oscillations in magnetic fields of galaxy clusters and the Milky Way on VHE $\gamma$-ray spectra is studied.
It is shown that this mechanism can lead to a substantial enhancement of the VHE flux and a spectral hardening, thus effectively reducing the opacity of the Universe to 
VHE $\gamma$-rays.
\end{abstract}

\section{Introduction}
Certain theories beyond the standard model predict the existence of pseudo-scalar pseudo Nambu-Goldstone bosons (or axion-like particles, ALPs) with a coupling to photons 
given by the Lagrangian
\begin{equation}
 \mathcal{L} = \frac{1}{4}g_{a\gamma}\tilde{F}_{\mu\nu}F^{\mu\nu}a = g_{a\gamma}\mathbf{E}\cdot\mathbf{B} a,
\label{eq:lagr}
\end{equation}
where $F^{\mu\nu}$ is the electromagnetic field tensor (with its dual $\tilde{F}$), $\mathbf{E}$ and $\mathbf{B}$ are the corresponding electric and magnetic fields, 
respectively, $a$ is the ALP field strength and $g_{a\gamma}$ is the coupling constant \cite[for a review]{jaeckel2010}.
The phenomenology is the same as for axions that were originally proposed to solve the strong CP problem in QCD \cite{pq1977,weinberg1978} 
with the difference that the photon-ALP coupling is unrelated to the ALP mass.
As a consequence of Eq. \ref{eq:lagr}, photons and ALPs mix in the presence of ambient magnetic fields transverse to the propagation direction of the photon beam.
The effect of these conversions on very high energy (VHE, energy $\gtrsim 100\,$GeV) $\gamma$-ray spectra of extragalactic sources has been studied in detail for different magnetic fields, 
e.g. the intergalactic magnetic field (IGMF) \cite{deangelis2011}, the field in the source \cite{sanchezconde2009}, or the $B$-field of the Milky Way \cite{simet2008}.
In general, two predictions can be made. First, a break in the energy spectra should be present at the energy for which the photon-ALP mixing becomes maximal 
(the so-called strong mixing regime) and, second, the mixing with ALPs can enhance the $\gamma$-ray flux.
The intrinsic VHE flux of an extragalactic source should be attenuated due to the interaction of $\gamma$-rays with low energy photons (ultra-violet to far infrared wavelengths)
of the extragalactic background light (EBL) \cite{hauser2001}.
The suppression is exponential and it scales with the optical depth $\tau(z,E)$, an increasing function with the energy $E$ of the $\gamma$-ray, the redshift $z$ of the source
and the EBL photon density.
Axion-like particles do not interact with EBL photons and can propagate unimpeded over cosmological distances. 
As a result, the VHE $\gamma$-ray flux can be enhanced in the optical thick regime (i.e. $\tau \gtrsim 1$), hardening the spectra of the sources.

Here, a scenario is studied in which the $\gamma$-ray source is located inside a galaxy cluster. 
The existence of magnetic fields in these clusters is well established \cite[for a review]{govoni2004} and thus photon-ALP conversions can occur.
Subsequently, the photon fraction of the beam is attenuated whereas ALPs can reconvert into photons in the galactic magnetic field (GMF) of the Milky Way.
Any photon-ALP conversion in the intergalactic magnetic field is neglected here
which is a reasonable assumption if field strengths are adopted that are preferred by large scale structure formations, i.e. $B_\mathrm{IGMF} \lesssim 10^{-11}$\,G \cite{dolag2005}.
The intra-cluster magnetic field (ICMF) and the GMF are modeled in the following way:
\begin{itemize}
 \item \emph{ICMF}. Observations of radio synchrotron and thermal X-ray emission together with Faraday rotation measures suggest random magnetic fields of the order of $\mu$G
 with coherence lengths of the order of 10\,kpc in galaxy clusters. Furthermore, the field strength follows the thermal gas distribution \cite{govoni2004}. 
 Here it is assumed that the $\gamma$-rays propagate through 100\,kpc of intra-cluster medium of constant density of $10^{-3}\mathrm{cm}^{-3}$ which is filled with an ICMF of constant field strength, $B_\mathrm{ICMF} = 1\,\mu\mathrm{G}$ with 
 a coherence length of 10\,kpc. A cell-like structure is adopted for the ICMF, i.e. the field strength is constant but the orientation changes randomly from one cell to another.
 \item \emph{GMF}. The regular component of the recent analytical model presented in \cite{jansson2012} is used here. It consist of a three components (disk, halo and a so-called X-component) and the 
 extension, orientation and strength of each component is determined by a fit to WMAP synchrotron polarization maps and rotation measures.
 Especially the latter two components result in a large photon-ALP conversion probability compared to other models.
 The random component of the GMF is not included here since the coherence length (which is typically assumed to be $\lesssim 100$\,pc) is much smaller than the photon-ALP oscillation length.
 The thermal electron density is assumed to be $10^{-2}\mathrm{cm}^{-3}$ and constant throughout the volume of the Milky Way.
\end{itemize}
The effects on the $\gamma$-ray flux will be studied with a photon-ALP coupling of $g_{a\gamma} = 5\times10^{-11}\mathrm{GeV}^{-1}$ which is below the current upper bound of the CAST experiment 
of $8.8\times10^{-11}\mathrm{GeV}^{-1}$ \cite{cast2007}.
A value of $m_a = 10$\,neV is adopted for the ALP mass so that the mixing entirely occurs in the strong mixing regime for the energies considered here.
Further details on the framework are provided in \cite{horns2012ICM}.

\section{Results}
\begin{figure}[t]
\centering
\includegraphics[width = 1\linewidth]{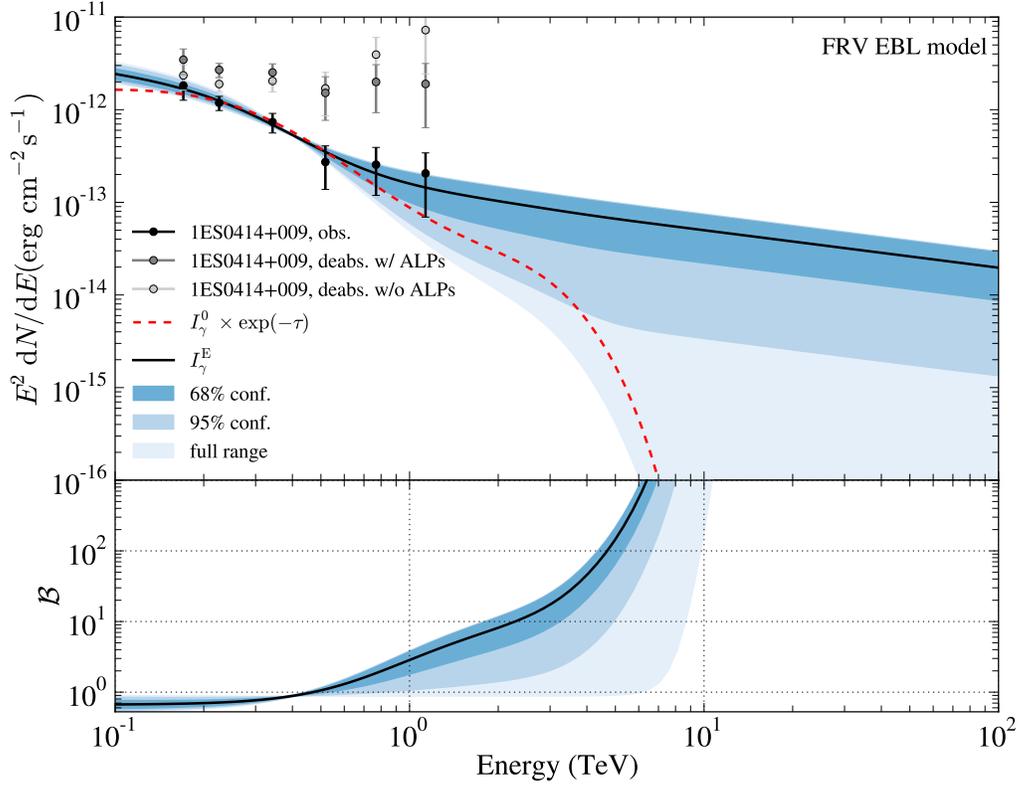} 
\caption{The effect of photon-ALP conversions on the spectrum of 1ES\,0414+009 measured with H.E.S.S. 
Top panel: The spectrum corrected for absorption with and without ALPs extrapolated to 100\,TeV.
Bottom panel: boost factor of the VHE $\gamma$-ray flux due to photon-ALP interactions.
}
\label{fig:spec}
\end{figure}

The observations conducted with the imaging air Cherenkov telescope array H.E.S.S.  of the blazar 1ES\,0414+009 \cite{1es0414hess2012} are used to study the effect of photon-ALP conversions on the VHE spectrum,
where the absorption due to the EBL is described with the model of \cite[FRV EBL below]{franceschini2008}.
This active galactic nuclei is believed to be located inside a galaxy cluster \cite[Tab. 1 and references therein]{horns2012ICM}. 
In total, 5000 realization of the random ICMF are simulated resulting in 5000 energy dependent photon survival probabilities $P_{\gamma\gamma}$. 
The mean of the distribution is used to correct the observed spectrum (black bullets in the top panel of Fig. \ref{fig:spec}). 
Clearly, the correction for the highest energy points of the spectrum is smaller with the ALP effect (dark gray bullets) than without it (light gray bullets) but they are still compatible 
within the statistical uncertainties (68\,\% confidence).
If the intrinsic spectrum $I_\gamma^0$ (fitted by a power law) extends up to 100\,TeV without an intrinsic cut off, ALPs should significantly increase the flux $I_\gamma^E$ observed on earth
(solid black line) whereas without ALPs the attenuation due to the EBL leads to a sharp cut off beyond $\sim2$\,TeV (red dashed line).
This is also eminent from the bottom panel of Fig. \ref{fig:spec} that displays the ratio between the two scenarios, $\mathcal{B} = P_{\gamma\gamma} / \exp(-\tau)$, the boost factor, which is
$\mathcal{B} \gtrsim 10$ above $\sim2$\,TeV.
However, the hardening of the observed spectrum depends on the actual orientation of the random magnetic fields and there exist realizations for which the attenuation 
is even stronger than without ALPs (light blue shaded area in Fig. \ref{fig:spec}). Nevertheless, the majority of realizations lead to significant boost of the flux.

\section{Conclusion}
The conversion of photons into ALPs in the magnetic fields of galaxy clusters and the Milky Way can lead to a substantial enhancement of the VHE $\gamma$-ray flux.
In contrast to the previously studied conversion in the intergalactic magnetic field for which only limits exist, the $B$-fields used here are all well established by observations.
The ALPs produced in galaxy clusters evade pair production with EBL photons and increase the transparency of the Universe for VHE $\gamma$-rays.
Thus, photon-ALP oscillation offer a mechanism beyond the standard model to explain indications found for a low opacity Universe inferred from VHE spectra \cite{horns2012}
and can be probed with the current and next generation of Cherenkov telescopes.

%\section*{Acknowledgments}
\paragraph{Acknowledgments}
MM would like to thank the state excellence cluster ``Connecting Particles with the Cosmos'' at the university of Hamburg.

% ****************************************************************************
% BIBLIOGRAPHY AREA
% ****************************************************************************

\begin{footnotesize}
\providecommand{\newblock}{}

\end{footnotesize}

% ****************************************************************************
% END OF BIBLIOGRAPHY AREA
% ****************************************************************************

\end{document}